\documentclass{pheauth}

\input epsf
\epsfverbosetrue

\begin{document}
\begin{frontmatter}

\title{The Hanbury Brown and Twiss Experiment with Fermions}

\author[address1]{S.~Oberholzer$^{1,}$},
\author[address1]{M.~Henny},
\author[address1]{C.~Strunk},
\author[address1]{C. Sch\"onenberger},
\author[address2]{T.~Heinzel},
\author[address2]{K.~Ensslin}, and
\author[address3]{M.~Holland}

\address[address1]{Institut f\"ur Physik, Universit\"at Basel,
Klingelbergstr.~82, CH-4056 Basel, Switzerland}

\address[address2]{Solid State Physics Laboratory, ETH Z\"urich,
CH-8093 Z\"urich, Switzerland}

\address[address3]{Department of Electronics, University of Glasgow,
Glasgow G12 8QQ, United Kingdom}

\thanks[thank1]{To whom correspondence should be addressed:
E-mail: stefan.oberholzer@unibas.ch}

\begin{abstract}
We realized an equivalent Hanbury Brown and Twiss experiment for a beam of
electrons in a two dimensional electron gas in the quantum Hall
regime. A metallic split gate serves as a tunable beam splitter which
is used to partition the incident beam into transmitted and
reflected partial beams. The current fluctuations in the reflected and transmitted
beam are fully anticorrelated demonstrating that fermions tend to exclude each other
(anti-bunching). If the occupation probability of the incident beam
is lowered by an additional gate, the anticorrelation  is reduced
and disappears in the classical limit of a highly diluted beam.
\end{abstract}

\end{frontmatter}
\begin{figure}[btp]
\begin{center}\leavevmode
\epsfxsize=76 mm \epsfbox{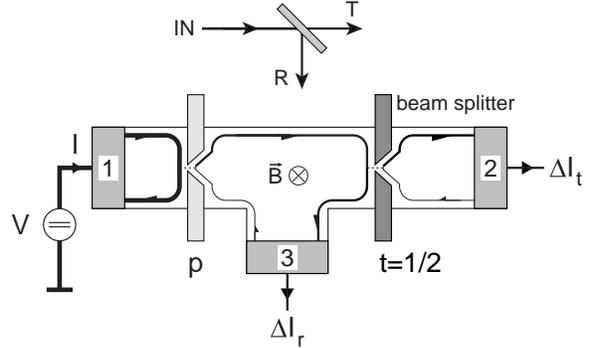} \vspace{-3mm}
\caption{Intensity correlation experiment for a degenerated beam
of electrons realized in a Hall bar connected to four electron
reservoirs (shaded). A metallic split gate serves as tunable beam
splitter. The  incident current can be depleted by an additional
gate $p$. Electrons escaping from \mbox{contact 1} travel along
the upper edge until reaching the beam splitter, where they are
either reflected into contact 3 or transmitted to contact 2. Note,
that the current incident to the beam splitter equals the current
$I$ flowing into \mbox{contact 1.}} \label{setup}
\end{center}
\end{figure}

The first quantum statistical measurements were carried out with photons
by Hanbury Brown and Twiss (HBT) in the 1950s. In a pioneering
experiment HBT determined the size of astronomical radio sources by measuring the
spatial coherence of the emitted radiation from correlations between
\emph{intensity} fluctuations at two different locations \cite{HBT54}.
In a subsequent optical table-top experiment,
they tested their idea by measuring intensity correlations for visible light:
the light beam of a Hg vapor
lamp (IN) was split into a transmitted (T) and a reflected beam (R) (see Fig.~\ref{setup}).
The equal-time intensity correlations between the two separated photon streams
was found to be positive (bunching) \cite{HBT56,Mo66}. This is a generic
property of particles obeying Bose-Einstein statistics.
In contrast to bosons, a negative correlation or `antibunching' is expected for
fermions, because fermions have to exclude each other due to the
Pauli principle.

In general, the time dependent number of particles $n(t)$ detected
during a certain time interval exhibits fluctuations
$\Delta n(t)=n(t)-\langle n\rangle$
around the average $\langle n\rangle$, called shot noise.
Shot noise in the electrical current is caused by
the discreteness of the electron charge and has been
extensively studied in submicrometer-sized nanostructures \cite{Bu90,DJ96}.
In a HBT type experiment, not the fluctuations in a single beam are
measured, but the correlation between fluctuations in the
transmitted (T) and reflected (R) beam originating from a beam splitter
(see Fig.~\ref{setup}). If the incident beam (IN) is not prepared at a
single-particle level, but shows intensity fluctuations $\Delta n$
around the mean value $\langle n\rangle$, the correlator $\langle
\Delta n_{t}\Delta  n_{r}\rangle$ between the fluctuations in the transmitted
$\Delta n_{t}$ and the reflected $\Delta n_{r}$ beam is given by:
\begin{equation}
    \langle\Delta n_{t}\Delta  n_{r}\rangle = t(1-t)\cdot
    \Bigl\{ \langle(\Delta n)^{2}\rangle-\langle n \rangle\Bigr\},
\label{cross}
\end{equation}
where $t$ is the transmission probability of the beam splitter.
The cross-correlation in Eq.~\ref{cross} is a sum of a term proportional to
$\langle(\Delta n)^{2}\rangle$, which depends on the particle statistics in the incident
beam, and a term proportional to  $\langle n \rangle$,
which is  caused by the probabilistic partitioning of single
particles at the beam splitter. This second term always gives a negative
contribution to the cross-correlation independent of whether the
particles are bosons or fermions.
The auto-correlation of the transmitted
(or reflected) beam is given by
\begin{equation}
    \langle(\Delta n_{t})^{2}\rangle = t(1-t)\cdot
    \Bigl\{ \frac{t}{1-t}\langle(\Delta n)^{2}\rangle+\langle n \rangle\Bigr\}.
\label{auto}
\end{equation}
Here, the sign of the second term is positive in
contrast to the cross-correlation in Eq.~\ref{cross}. For an incident
beam of particles obeying \emph{Poisson} statistics with
$\langle(\Delta n)^{2}\rangle=\langle n \rangle$ the cross-correlation is zero (see
Eq.~\ref{cross}). For \emph{super-Poisson} statistics,
$\langle(\Delta n)^{2}\rangle>\langle n \rangle$, the cross-correlation
is positive (HBT result for thermal light). In contrast, anticorrelation
results if the noise in the incident beam is \emph{sub-Poissonian}
$\langle(\Delta n)^{2}\rangle<\langle n \rangle$. Maximal
anticorrelation is obtained if the incident beam carries no
fluctuations ($\langle(\Delta n)^{2}\rangle=0$), which is the case for
a completely degenerated electron beam at zero temperature. Noise suppression
in a single beam according to the Pauli principle
has recently been found in electrical
measurements on quantum-point contacts and nanowires \cite{Re96}.

A HBT-type intensity correlation experiment
for particles with non classical Fermi-Dirac statistics has not yet been carried out.
Attempts to measure the expected negative correlation in
a beam of free electrons have not been successful yet, mainly because
of the low particle density in such a beam.
The HBT experiment for fermions has recently
been carried out in two independent experiments based on semiconductor
devices \cite{Ob99}. Theoretically, these experiments
and other multiterminal correlation experiments have been considered
before \cite{Bu90,Ma92,Bu92,Ya92}.

\begin{figure}[btp]
\begin{center}\leavevmode
\epsfxsize=76 mm \epsfbox{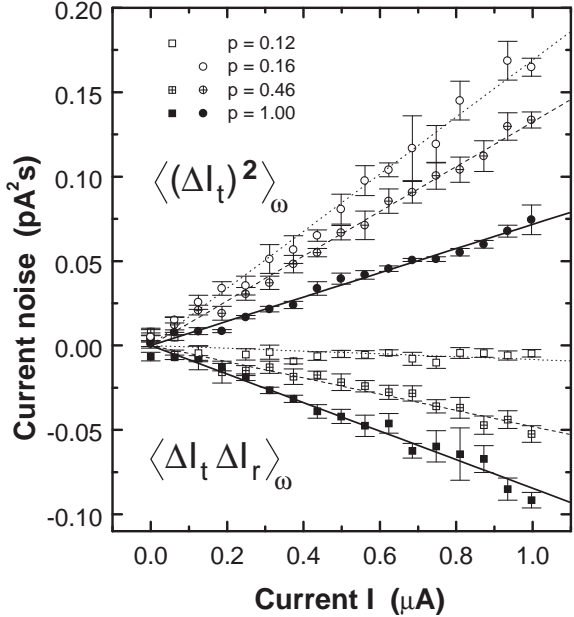} \vspace{-2mm}
\caption{Measured spectral densities of correlations between
current-fluctuations as a function of the current $I$ of the
incident beam at $T=2.5$ K. $\langle \Delta I_{t}\Delta
I_{r}\rangle_{\omega}$ denotes the cross-correlation between the
transmitted and reflected beams and $\langle (\Delta
I_{t})^{2}\rangle_{\omega}$ the auto-correlation in the
transmitted beam. The offset noise arising from thermal
fluctuations and residual amplifier noise has been subtracted. The
absolute slopes are $(0.23\pm 0.030)\cdot 2e$ and $(0.26\pm
0.037)\cdot 2e$ for the auto-correlation and cross-correlation,
respectively, in agreement with the expected prefactor
$t(1-t)=1/4$. If the incident beam is diluted by lowering the
transmission probability $p$, the cross-correlation gets smaller
and disappears for $p$ close to zero. The auto-correlation itself
increases, because the incident beam is noisy.} \label{noise}
\end{center}
\end{figure}

One way to realize a HBT experiment with electrons has been proposed
by B\"uttiker \cite{Bu90}: In a two dimensional electron gas (2 DEG)
in the quantum Hall regime the current flows in one-dimensional channels along
the edges of the device (see Fig.~\ref{setup}). These edge channels can be
used to separate the incident from the reflected beam.
If there would be no magnetic
field the current would also flow in the bulk and incident and
reflected beam could not be distinguished.
The beam splitter is a lithographically patterned metallic split gate,
which can be tuned by an applied voltage.
In addition, there is a second gate with transmission probability $p$ before the beam
splitter which is used to dilute the occupation in the incident beam.
Applying a constant voltage $V$ to contact 1 the charge current $I$ is
injected into the Hall bar. The magnetic field perpendicular to the 2
DEG is adjusted to filling factor $\nu = 2$, so that the current flows in
one spin-degenerated edge state. Let us first consider the case,
where the additional gate $p$ is not
used: the electrons escaping from contact 1 travel along the upper edge until reaching the
beam splitter, where they are either transmitted with probability $t$ to
leave the device at contact 2, or reflected with probability $r=1-t$ leaving at contact 3.
Provided $eV\gg kT$, the theory predicts for the spectral densities of the auto- and
cross-correlation according to Eq.~\ref{cross} and \ref{auto} \cite{Bu92}
\begin{equation}
    \langle \Delta I_{\alpha}\Delta I_{\beta}\rangle_{\omega}=\pm 2e |I|t(1-t)
    \label{b}
\end{equation}
with $e$ the electron charge and $\alpha$, $\beta$ either $t$ or $r$.
The positive sign corresponds to the auto-correlation, where
$\alpha=\beta$, and the negative one to the cross-correlation with
$\alpha\neq\beta$. Because the cross-/auto-correlation is largest for
$t=1/2$ the beam splitter is adjusted
to transmit and reflect electrons with 50 $\%$ probability.

Figure~\ref{noise} shows the cross-correlation
$\langle\Delta I_{t}\Delta I_{r}\rangle_{\omega}$ (solid squares) of the
fluctuations $\Delta I_{t}$ and $\Delta I_{r}$ versus bias current
$I$ at $T=2.5$ K. A nearly linear dependence with a negative slope is
found showing that the fluctuations are indeed anticorrelated.
The auto-correlation (solid circles) of the transmitted current
$\langle(\Delta I_{t})^{2}\rangle_{\omega}$ (or reflected
current, not shown) has a positive slope. The negative
cross-correlation and the positive auto-correlation are equal in magnitude confirming that
the partial beams are fully anticorrelated. We can therefore conclude
that there is no uncertainty in the occupation of the
incident beam, that is $\langle(\Delta n)^{2}\rangle = 0$ in Eq.~\ref{cross} and
\ref{auto}. All states in the incident beam are occupied with probability one and
hence are noiseless by virtue of the Pauli principle. Formally, this
follows also from $\langle(\Delta n)^{2}\rangle =
\langle (\Delta n_{t}+\Delta n_{r})^{2}\rangle =
\langle(\Delta n_{t})^{2}\rangle+2\langle\Delta n_{t}\Delta n_{t}\rangle+
\langle(\Delta n_{r})^{2}\rangle=0$ within experimental accuracy.
The fact that the current $I$ of the incident beam is noiseless demonstrates that the
constant voltage applied to reservoir 1 is converted into a \emph{constant}
current $e^{2}V/h$ (per accessible mode) according to the
fundamental requirement of the Landauer-B\"uttiker formalism.

\begin{figure}[hbtp]
\begin{center}\leavevmode
\epsfxsize=80 mm \epsfbox{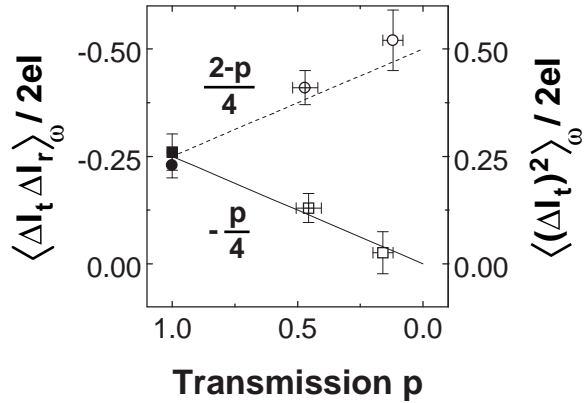} \vspace{-2mm}
\caption{Dependence of the auto- and cross-correlation on the
transmission probability $p$ of the additional gate for $t=$ 1/2.
If $p$ is lowered from $1$ to $0$ the statistics in the incident
beam are changed from Fermi-Dirac to Poisson statistics. The
normalized auto-correlation $\langle (\Delta
I_{t})^{2}\rangle_{\omega}/2eI$ increases from $1/4$ to $1/2$ and
the normalized cross-correlation $\langle \Delta I_{t} \Delta
I_{r}\rangle_{\omega}/2eI$ from $-1/4$ to $0$ in good agreement
with the prediction of Eq.~\ref{auto2} and \ref{cross2}.}
\label{noise2}\end{center}\end{figure}

In an extension of our experiment we have changed the statistics in
the incident beam using the additional gate with transmission
$p\in[0,1]$. If  $\langle n \rangle$ is the mean particle number of the
beam incident to the beam splitter, the particle number behind the  first gate
$p$ is given by
$\langle\tilde{n}\rangle=p\langle n\rangle$ with noise
$\langle(\Delta\tilde{n})^{2}\rangle=p(1-p)\langle n\rangle$.
Using Eq.~\ref{cross} and \ref{auto} the normalized auto- and cross-correlation
depend on the transmission probability $p$ as
\begin{eqnarray}
    \frac{\langle(\Delta I_{t})^{2}\rangle_{\omega}}{2eI} & =
    & t(1-pt) = \frac{2-p}{4}\label{auto2}  \\
    \frac{\langle\Delta I_{t}\Delta I_{r}\rangle_{\omega}}{2eI} & =
    & -t(1-t)p = -\frac{p}{4},\label{cross2}
\end{eqnarray}
for $t=$ 1/2.
If $p$ decreases from 1 to 0 the states in the
incident beam are diluted and the anticorrelation (Eq.~\ref{cross2}) becomes
 smaller (`+-centered' and
open squares in Fig.~\ref{noise}). In case of very low transmission $p$ the
statistics in the incident beam is Poissonian and the anticorrelation
disappears as discussed above. The auto-correlation itself increases
(`+-centered' and open circles in Fig.~\ref{noise}),
because of the noise in the incident beam (see Eq.~\ref{auto}).
The dependence of the auto- and
cross-correlation on the probability $p$ is shown in Fig.~\ref{noise2}.
The measured slopes of the data in Fig.~\ref{noise} are in good
agreement with the predictions of Eq.~\ref{auto2} and \ref{cross2}
within experimental accuracy.

In conclusion, a HBT type experiment with electrons has been realized
in a solid state device. We demonstrate that current correlations are
sensitive to the particle statistics in the incident beam. Full anticorrelation
is observed for electrons obeying Fermi-Dirac statistics,
whereas the anticorrelation is gradually suppressed if the incident
beam is diluted. It would be interesting
to extend this kind of experiments to electronic states
obeying different statistics like the
fractional quantum Hall state \cite{Gl97}. The observation of bunching of
electrons might be possible by preparing
electronic states with fluctations larger than the classical Poisson
value, which have recently been observed in resonant tunneling
devices and in superconducting weak links \cite{In98}.

The authors would like to thank M.~B\"uttiker, D.~C.~Glattli, and
T.~Gramespacher for valuable comments. This work was supported by the
Swiss National Science Foundation.


\end{document}